\documentclass{mem}
\usepackage{natbib}\usepackage{txfonts}\usepackage{balance}
\usepackage{graphicx}
\usepackage[a4paper,breaklinks,dvipdfm]{hyperref}
\idline{000}{001}
\begin{document}
\def\teff{$T\rm_{eff }$}
\def\kms{$\mathrm {km s}^{-1}$}

\title{
Varying constants and dark energy with the
E-ELT
}

   \subtitle{}

\author{
P. \,Vielzeuf\inst{1,2}  \and  C.\ J.\ A.\ P.\ Martins\inst{1}
          }

  \offprints{P. E. Vielzeuf}

\institute{
Centro de Astrof\'{\i}sica da Universidade do Porto, Rua das Estrelas, 4150-762 Porto, Portugal
\and
Faculdade de Ci\^encias, Universidade do Porto, Rua do Campo Alegre 687, 4169-007 Porto, Portugal
\email{Pauline.Vielzeuf@astro.up.pt}
}

\authorrunning{Vielzeuf \& Martins}

\titlerunning{Varying constants \& dark energy with the E-ELT}

\abstract{
The observational possibilities enabled by an ultra-stable CODEX-like spectrograph at the E-ELT will open new doors in the characterisation of the nature of Dark Energy. Indeed, it will provide measurements of a so far unexplored redshift-range ($2<z<5$) and will carry out simultaneously cosmological tests such as the Sandage-Loeb test and precision tests of the standard model. Here we will illustrate how, with these abilities, such spectrographs---alone or in synergy with other facilities---will manage to constrain cosmological scenarios and test classes of models that would otherwise be difficult to distinguish from the standard $\Lambda{\rm CDM}$ paradigm.
}
\maketitle{}

\section{Introduction}

The recent observation of the universe's cosmic acceleration demonstrates that the standard model of cosmology ($\Lambda$CDM) is incomplete or incorrect. Indeed, $70\%$ of the present universe density remains uncharacterised because unobserved---the so-called Dark Energy. Therefore, it seems natural to wonder if there is a new kind of physics behind this Dark Energy.
\par 
 On the other hand, recent observations suggested variations of the fine-structure constant $\alpha=\frac{e^2}{\hbar c}$ \citep{webb}. But this measurements are not precise enough to be accepted. With future facilities such as HIRES an Ultra‐stable high-resolution spectrograph for the E-ELT, or ESPRESSO for the VLT, one will significantly improve the precision on the variation of fundamental constants (reaching $\sigma_{\Delta\alpha}\sim {\rm few}\times 10^{-8}$ for HIRES). \par
Even more exciting than testing the stability of fundamental constants, this future E-ELT spectrograph will also perform a first direct measurement of  
the accelerating expansion of the  universe by observing the drift of quasar absorptions lines through the Lyman-$\alpha$ forest. Finally, it will make measurements of the CMB temperature at redshifts $z>0$, a key consistency test. The accuracy expected in this temperature variations are:
\begin{equation}
\Delta T_{ESPRESSO}\sim0.35K
\end{equation}
and,
\begin{equation}
\Delta T_{HIRES}\sim 0.07K
\end{equation}

\section{Varying fundamental constant and Dark energy}
If the fundamental constants of nature vary, one can treat the problem in two distinct ways:
\begin{enumerate}
\item Dark energy and varying constants are due to the same additional degree of freedom.
\item The variation of $\alpha$ is due to some other field with negligible contribution to the universe energy density. \end{enumerate}
The second case will be explored through a specific class of model later in these proceedings. For the moment, let's assume the the presumed scalar field giving dark energy couples to the electromagnetism and induces a variation of fundamental constants. Then one can derive the equation of state as a function of the variation of fundamental constants. \par 
Two sub-cases can now be considered:

\subsection{$w\ge-1$}

The dark energy equation of state of this kind of models can be expressed as:
\begin{equation}\label{eq:eosfield}
w=\frac{P_\Phi}{\rho_\Phi}=\frac{\dot\Phi^2-2V(\Phi)}{\dot\Phi^2+2V(\Phi)},
\end{equation}
and it can be reconstructed as \citep{nunes}
\begin{equation}\label{eq:(w+1)n2}
w+1=\frac{(\kappa\Phi')^2}{3\Omega_\Phi},
\end{equation}
where $\kappa^2=8\pi G$, $\Omega_\Phi$ is the dark energy density parameter and the primes denote derivatives with respect to $N=\ln(a)$ (a being the scale factor).
\par
Finally, it can be related to $\alpha$ as:
\begin{equation}\label{phivaralphamu}
\Phi'=\frac{\alpha'}{\kappa\zeta\alpha}
\end{equation}
$\zeta$ being the strength of the coupling.

\subsection{Phantom case : $w< -1$}

In this case one can write the phantom field energy density and its pressure as \citep{Singh}
\begin{equation}
\rho_\Phi=-\frac{\dot\Phi^2}{2}+V(\Phi),
\end{equation}
\begin{equation}
P_\Phi=-\frac{\dot\Phi^2}{2}-V(\Phi)
\end{equation}
Hence we have
\begin{equation}
P_\Phi+\rho_\Phi=-\dot\Phi^2\Rightarrow P_\Phi=-\dot\Phi^2-\rho_\Phi
\end{equation}
As in a case of $w+1\le0$, one can rewrite the equation of state as: 
\begin{equation}
w=\frac{-\dot\Phi^2-\rho_\Phi}{\rho_\Phi}=-1-\frac{\dot\Phi^2}{\rho_\Phi}.
\end{equation}
Then knowing that by definition 
\begin{equation}
\rho_\Phi=\frac{3H^2\Omega_{\Phi}}{\kappa^2},
\end{equation}
and that,
\begin{equation}
\dot\Phi^2=\left(\frac{d\Phi}{dt}\right)^2=\left(\frac{d\Phi}{dN}\right)^2\left(\frac{dN}{dt}\right)^2=\Phi'^2H^2,
\end{equation}
the equation of state for the phantom case can be written as: 
\begin{equation}
w+1=-\frac{\kappa^2\dot\Phi^2}{3H^2\Omega_\Phi}=-\frac{(\kappa\Phi')^2}{3\Omega_\Phi}.
\end{equation}
Substituting then equation (\ref{phivaralphamu}), one obtains:
\begin{equation}
w+1=-\frac{(\frac{\alpha'}{\alpha})^2}{3\zeta_\alpha^2\Omega_\Phi}.
\end{equation}

\section{Sandage-Loeb test}
The Sandage-Loeb test \citep{sandage,loeb}, provides a way to distinguish cosmological models by comparing their expansion rate. Indeed, the evolution of the Hubble expansion causes the redshifts  
of distant objects to change slowly with time. Starting from the definition of the redshift, one can derive a relation between the redshift-drift and the Friedmann equation of the considered model. \par 
\begin{equation}
\Delta z=\Delta t_0\times H_0\times\left[1+z_s-\frac{H(z)}{H_0}\right]
\end{equation}
Where $z_s$ is the source redshift, $t_0$ the time of observation, and $H(z)$ and $H_0$ the Hubble parameter at redshift z and today.\par
One observes spectroscopic velocities, which are related to the redshift drift via 
\begin{equation}
\Delta v=c\times \frac{\Delta z_s}{(1+z_s)}
\end{equation}
One can then insert the Friedmann equation of the considered class of model to infer the corresponding cosmological parameters; for example, the general Friedmann equation for a varying equation of state of dark energy can be expressed as:
\begin{equation}
\left(\frac{H}{H_0}\right)^2=\frac{\Omega_m}{a^3}+\frac{\Omega_\gamma}{a^4}+\frac{\Omega_k}{a^4}+\frac{\Omega_\Lambda}{e^{3\int\frac{da(1+w(a))}{a}}},
\end{equation}
where the $\Omega_i$ represents respectively the density parameters of matter, radiation, curvature and dark energy today.

\section{CMB temperature test}

Another test to be carry out by HIRES and other facilities is measuring the CMB temperature at redshift $z>0$. In the standard model, the CMB temperature evolves adiabatically as
\begin{equation}\label{eq:adiabtemp}
T=T_0(1+z)
\end{equation}
$T_0$ being the CMB temperature today.
This relation can be violated if photons couple to scalar or pseudo-scalar degrees of freedom.
One can then define a correction to the standard case $y(z)$ such that:
\begin{equation}
T=T_0(1+z)y(z)
\end{equation}
$y(z)=1$ corresponding to the standard case.\par
 A simple parametrisation for this deviation from the standard case can be expressed as
\begin{equation}
T=T_0(1+z)^{1-\beta}
\end{equation}
Current constraints on $\beta$ are relatively weak, but will be significantly improved by the E-ELT \citep{Avgoustidis}.

\section{Cosmological case studies}

\subsection{Early Dark Energy}

This model proposed in \citep{ede} suggests that the dark energy remains a significant fraction of the universe’s energy density at all times.
\par
The relations parametrising dark energy for this class of model are:
\begin{equation}
\Omega_{de}(a)=\frac{\Omega_{de}^{0}-\Omega_e(1-a^{-3w_0})}{\Omega_{de}^{0}+\Omega_m^0a^{3w_0}}+\Omega_e(1-a^{-3w_0})
\end{equation}
\begin{equation}
w(a)=-\frac{1}{3(1-\Omega_{de}(a))}\frac{{\it d}ln\Omega_{de}(a)}{{\it d}ln(a)}+\frac{a_{eq}}{3(a+a_{eq})}
\end{equation}
where $\Omega_{e}$ represents the amount of early dark energy, $\Omega_{de}^0$ the amount of dark energy today, $a$ the scale factor, $a_{eq}$ the scale factor at matter-radiation equality and $w_0$ the equation of state of dark energy today.  

\begin{figure}[h]
\includegraphics[width=7cm]{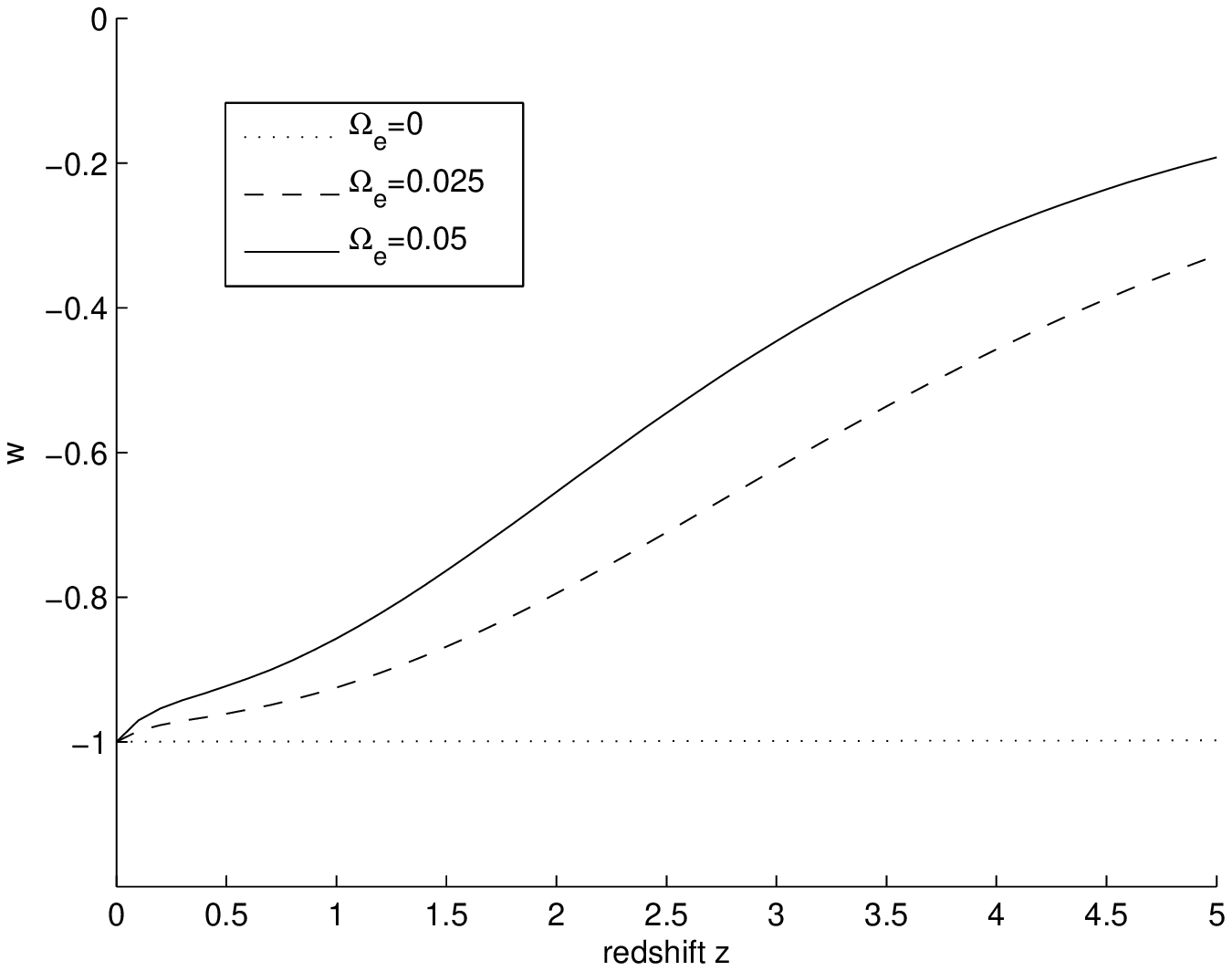}
\includegraphics[width=7cm]{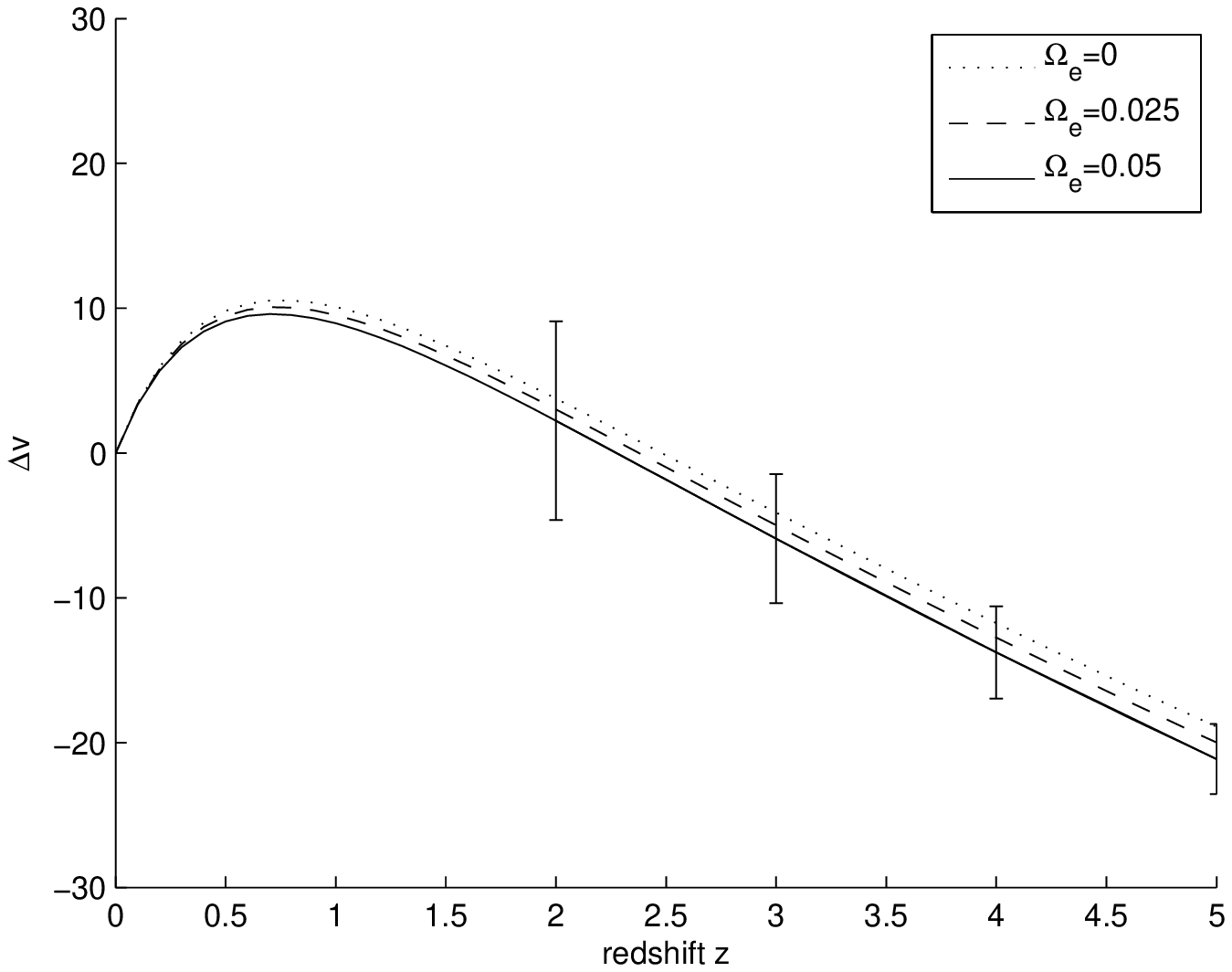}
\caption{Equation of state (top) and SL signal (bottom) for different values of the parameter $\Omega_e$,
and for a SL observation time of $\Delta t = 20$ years, with the vertical bars being the
HIRES measurement accuracy expected.}
\label{edeesosl}
\end{figure}

The equation of state and the Sandage-Loeb signal of this class of model has been plotted in Fig. \ref{edeesosl} for different amounts of Early Dark energy.

\begin{figure}[h]
\includegraphics[width=7cm]{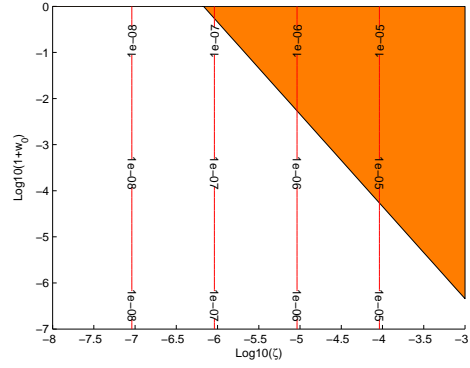}
\caption{
\footnotesize The relative variation of the fine-structure constant, $\frac{\Delta\alpha}{\alpha}$, at
redshift $z=4$, as a function of $\zeta$ and $w_0$ , with $\Omega_e = 0.05$. The shaded region is the local atomic bound $\zeta\sqrt{3\Omega_{\Phi,0}(1+w_0)}<10^{-6}$}
\label{edealphavar}
\end{figure}

As one can see on the figure, even though the equation of state varies significantly with the amount of early dark energy chosen, HIRES won't be able (through the Sandage-Loeb test alone) to discriminate this class of model from $\Lambda CDM$. The next step is then to verify if it will be able to constrain it through the variation of fundamental constants. 
Assuming an amount of early dark energy of $\Omega_e = 0.05$, and using Eqs. (\ref{eq:(w+1)n2}--\ref{phivaralphamu}) one can predict the relative variation of the fine-structure constant at $z=4$, as shown in Fig. \ref{edealphavar}.

Given the expected accuracy of HIRES mentioned earlier, one will be able to detect such variations; this highlight the importance of having a spectrograph capable of doing both measurements.
\subsection{BSBM}
In the Bekenstein-Sandvik-Barrow-Magueijo model \citep{BSBM}
the dark energy is due to a cosmological constant, while
the variation of $\alpha$ is due to some other field with negligible contribution to the 
universe energy density and can be parametrised as
\begin{equation}
\frac{\Delta\alpha}{\alpha}=-4\epsilon{\rm ln}(1+z)
\end{equation} 
where $\epsilon$ gives the magnitude of the variation.\par 
If one wrongly assumes that the dark energy is due to the $\alpha$-field, one reconstructs the equation of state \citep{nunes}: 
 \par 
\begin{equation}
 w(N)=(\lambda^2-3)\left[3-\frac{\lambda^2}{w_0}\frac{\Omega_{m,0}}{\Omega_{\Phi,0}}e^{(\lambda^2-3)N}\right]^{-1}
\end{equation}
 with $\lambda=\sqrt{3\Omega_{\Phi,0}(1+w_0)}=4\frac{\epsilon}{\zeta}$.\par
\begin{figure}[h]
\includegraphics[width=7cm]{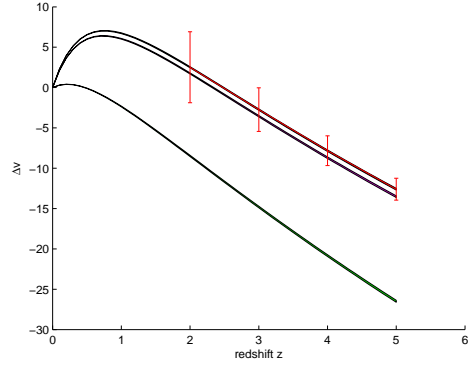}
\caption{
\footnotesize The SL test for reconstructed BSBM models 
with $\lambda=1$ (bottom band) and $\lambda=0.3$, compared 
to the standard $\Lambda$CDM case (top band). The 
bands correspond to the range of 
$\Omega_{\Phi,0}=0.73\pm0.01$.}
\label{bsbmsl}
\end{figure}

The corresponding Sandage-Loeb signal for different values of the parameter $\lambda$ is shown in Fig. \ref{bsbmsl}.

For this case,as one can see on the figure, the Sandage-Loeb test will be efficient only for large values of $\lambda$, but small $\lambda$ meaning large coupling one expect to discriminate this range of $\lambda$ with future laboratory tests. As a result, this wrong assumption can always be identified.\par

\begin{figure}[h]
\includegraphics[width=7cm]{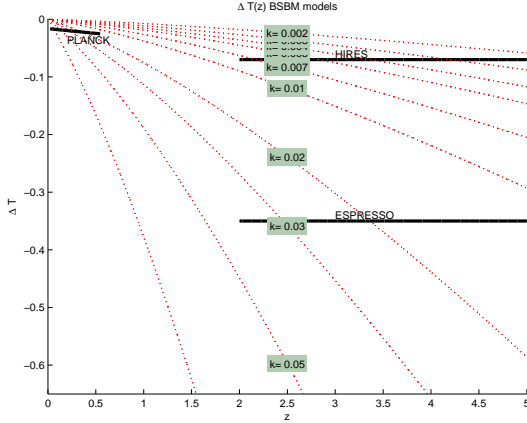}
\caption{
\footnotesize Variation of the temperature (relative to the standard
model) as function of $z$ in a BSBM class of models,
for different values of k and using $T_0 = 2.725 \pm 0.002$. Also
depicted are the limits of detection of this difference with
HIRES, ESPRESSO and Planck clusters. The span of each
bar is meant to represent the redshift range of each set of
measurements.}
\label{bsbmtempvar}
\end{figure}

Looking now at the CMB temperature-redshift relation for this class of model, as it has been done in \citep{moi4}, one gets the following:

\begin{equation}
T(z)=T_0(1+z)[1-k{\rm ln}(1+z)]
\end{equation}
One can then predict what would be the difference in temperature between the standard case of Eq. (\ref{eq:adiabtemp}) and the BSBM for different values of the coupling $k$. This difference is illustrated in Fig. \ref{bsbmtempvar}.
 
The figure shows well that it won't be possible to detect a variation of $T(z)$ at the order that the variation have been detected without having a large number of sources which are not known today.

\section{Complementary with other facilities}


\begin{table*}[t]
\caption{$1-\sigma$ ($2-\sigma$) uncertainty in the relevant model parameters, marginalising over the others. {\it Weak} and {\it Strong} correspond to different priors for current data.}
\label{table1}
\centering
\resizebox{\linewidth}{!}{
\begin{tabular}{ccccc}
\hline\hline                       
Dataset & $\delta w_0$ & $\delta w_a$ & $\delta\Omega_m$ & $\delta k$ \\ [0.5ex]
\hline                
Current (weak) & 0.25 & 1.3 & 0.06 & $1.2\times 10^{-6} (10^{-5})$ \\  
Current (strong) & 0.22 & 0.65 & 0.06 &  $1.2\times 10^{-6} (10^{-5})$\\ 
Euclid (BAO)+ SNAP & 0.15 (0.35) & 0.4 (1.6)  & 0.03 &$ 10^{-6} (1.1\times 10^{-6})$ \\ 
Euclid only (BAO+SN) & 0.15 (0.35) & 0.6 (1.6) & 0.03 &- \\ 
Euclid (BAO+SN)+SNAP & 0.14 (0.35) & 0.8 (1.5) & 0.025 &$8\times 10^{-7} (9\times 10^{-6})$ \\
Euclid (BAO)+SNAP+E-ELT & 0.13(0.30)& 0.755(1.45)&0.023 &$8\times 10^{-7} (9\times 10^{-6})$ \\
Euclid (BAO)+SNAP+TMT & 0.13 (0.25)& 0.4 (1.3) &0.024 &$6\times 10^{-7} (8\times 10^{-6})$ \\
  \end{tabular}
}
\end{table*}


It's also pertinent to verify whether or not one can break degeneracies in the parameter space by combining HIRES results with other facilities. This has been done in \citep{moi3} in the context of Planck.

\begin{figure}[h]
\includegraphics[width=7cm]{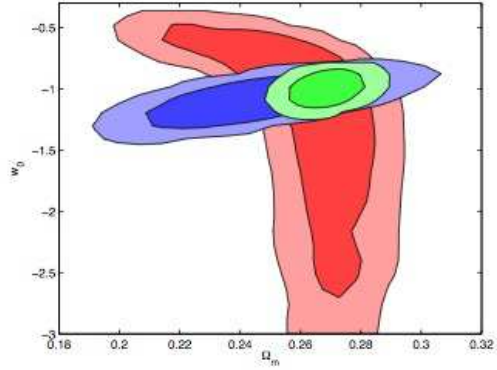}
\caption{
\footnotesize 2-D constraints on $w_0$ and $\Omega_m$ using CMB (blue), SL (red) and combining the two probes (green).}
\label{complementarity}
\end{figure}


Fig. \ref{complementarity} illustrates how degeneracies can be broken by combining Sandage-Loeb and CMB results when estimating cosmological parameters. The crucial role of the redshift drift is not its intrinsic sensitivity, but the fact that it is sensitive to parameters that are not easily measured by other probes.

Another study illustrates the improvements on the constraints on varying fundamental constants by combining probes of several facilities \citep{moi4}.
In this work we assumed a CPL parametrisation for the equation of state of dark energy
\begin{equation}
w(a)=w_0+w_a(1-a)
\end{equation}

Considering then that the field coupling to radiation is the one responsible for dark energy, one obtains the following CMB temperature evolution equation 

\begin{equation}
y^4(a)=1+3k\frac{\Omega_{\Phi,0}}{\Omega_{\gamma,0}}\int^a_1x^{-3(w_0+w_a+k)}e^{-3w_a(1-x)}dx
\,.
\end{equation}

By integrating this equation numerically, one finds (as in the BSBM case) the temperature difference between the standard model and this class of models and given the accuracy expected for future facilities mentioned before, one found that one will get enough precision to put stronger bounds on these classes of models \citep{moi4}.\par

Forecasts of future constraints were obtained for simulated datasets for Euclid, SNAP, TMT and E-ELT, for a CPL parametrisation allowing for photon number non-conservation. Supernovas for Euclid and SNAP won't exceed a redshift of $z\sim 1.7$, however with JWST support the TMT expects to find about 250 supernovas in the redshift range $1<z<3$, while for the E-ELT one expects 50 supernovas in the range $1<z<5$. Using this one forecast the constraints that one can reach on $w_0$, $w_a$ and the coupling $k$, the results are shown in Table \ref{table1}.

The results show in a first time that Euclid on its own will be able to constrain dark energy even when allowing for the photon number non-conservation. One can also notice that high-z supernovas improve noticeably the constraints.

\section{Conclusions}

We illustrated the abilities of HIRES to probe the nature of Dark Energy in the otherwise unexplored redshift range $2<z<5$.The results also show that being able to simultaneously carry out the SL test and precision tests of the standard model  gives HIRES a unique advantage over other spectrographs. \par

We also highlighted how Sandage-Loeb observations alongside CMB data can break degeneracies between different parameters.
 Euclid will be able on its own to provide constraints on Dark Energy parameters while allowing for photon 'non-conservation'. Stronger constraints can be provided by combining the probes especially with high-z SN from the E-ELT.

\begin{acknowledgements}
We acknowledge the financial support of grant PTDC/FIS/111725/2009 from FCT (Portugal). CJM is also supported by an FCT Research Professorship, contract reference IF/00064/2012.
\end{acknowledgements}

\bibliographystyle{aa}

\end{document}